\documentclass[10pt,twocolumn,twoside]{IEEEtran}

\def\argmin{\mathop{\rm arg\,min}}
\usepackage{graphicx,epsfig}
\usepackage{cite}
\usepackage{stfloats}
\usepackage{amsmath}
\usepackage{amsopn}
\usepackage{amsfonts,helvet}
\usepackage{caption}
\usepackage{subcaption}
\usepackage{flushend}
\usepackage{subfig}
\usepackage{hhline}
\usepackage{amsmath}

\newtheorem{theorem}{Theorem}

\newtheorem{algo}[theorem]{Algorithm}

\newtheorem{proposition}[theorem]{Proposition}

\newcounter{proposition}
\begin{document}

\title{A Universal Channel Model for Molecular Communication Systems with Metal-Oxide Detectors}

\author{\authorblockN{Na-Rae Kim,$^{\ast \ddag}$ Nariman Farsad,$^{\dag \ddag}$  Chan-Byoung Chae,$^\ast$ and Andrew W. Eckford$^\dag$}\\
\authorblockA{$^\ast$School of Integrated Technology, Yonsei Institute of Convergence Technology, Yonsei University, Korea\\
Email: \{nrkim, cbchae\}@yonsei.ac.kr\\
$^\dag$Department of Electrical Engineering and Computer Science, York University, Canada\\
Email: \{nariman, aeckford\}@cse.yorku.ca\\}
\thanks{\ddag The authors have equally contributed.}}

\maketitle \setcounter{page}{1} 

\begin{abstract}
In this paper, we propose an end-to-end channel model for molecular communication systems with metal-oxide sensors. In particular, we focus on the recently developed table top molecular communication platform. 
The system is separated into two parts: the propagation and the sensor detection. There is derived, based on this, a more realistic end-to-end channel model. However, since some of the coefficients in the derived models are unknown, we collect a great deal of experimental data to estimate these coefficients and evaluate how they change with respect to the different system parameters. Finally, a noise model is derived for the system to complete an end-to-end system model for the tabletop platform. 
\end{abstract}

\begin{keywords}
Molecular communication, channel model, tabletop platform, metal-oxide sensor.
\end{keywords}

\section{Introduction}
\label{Sec:Intro}

In molecular communication, information is transferred via chemical signals. Unlike the traditional wireless communication systems, molecules are used to physically carry information~\cite{eckBook}. The transmitter releases molecules, often called messenger/information molecules, that propagate through diffusion~\cite{nak08, pie10, pie13}, medium flow~\cite{sri12, far12NanoBio}, or active transport~\cite{hiy10LabChip, far12NANO, far14TSP} to arrive at the receiver. The information can be encoded in several ways. For example, 
different information can be represented by different concentration, number, and/or type of molecules~\cite{mah10,far12NanoBio,cob10}.  

In the literature, most prior work has focused on the communication via diffusion, or diffusion assisted by flow. Most of these works, moreover,  are only theoretical-based analyses of these systems. The authors in~\cite{kim13} analyzed the maximum data transmission rates of diffusion-based systems. There has also been several work regarding proper modulation techniques for molecular communication~\cite{kim13}. Some potential applications have been addressed in~\cite{Akyildiz_cn08}, and one of the promising applications for molecular communication is considered to be the biomedical field.

\begin{figure}[t]
 \centerline{\resizebox{\columnwidth}{!}{\includegraphics{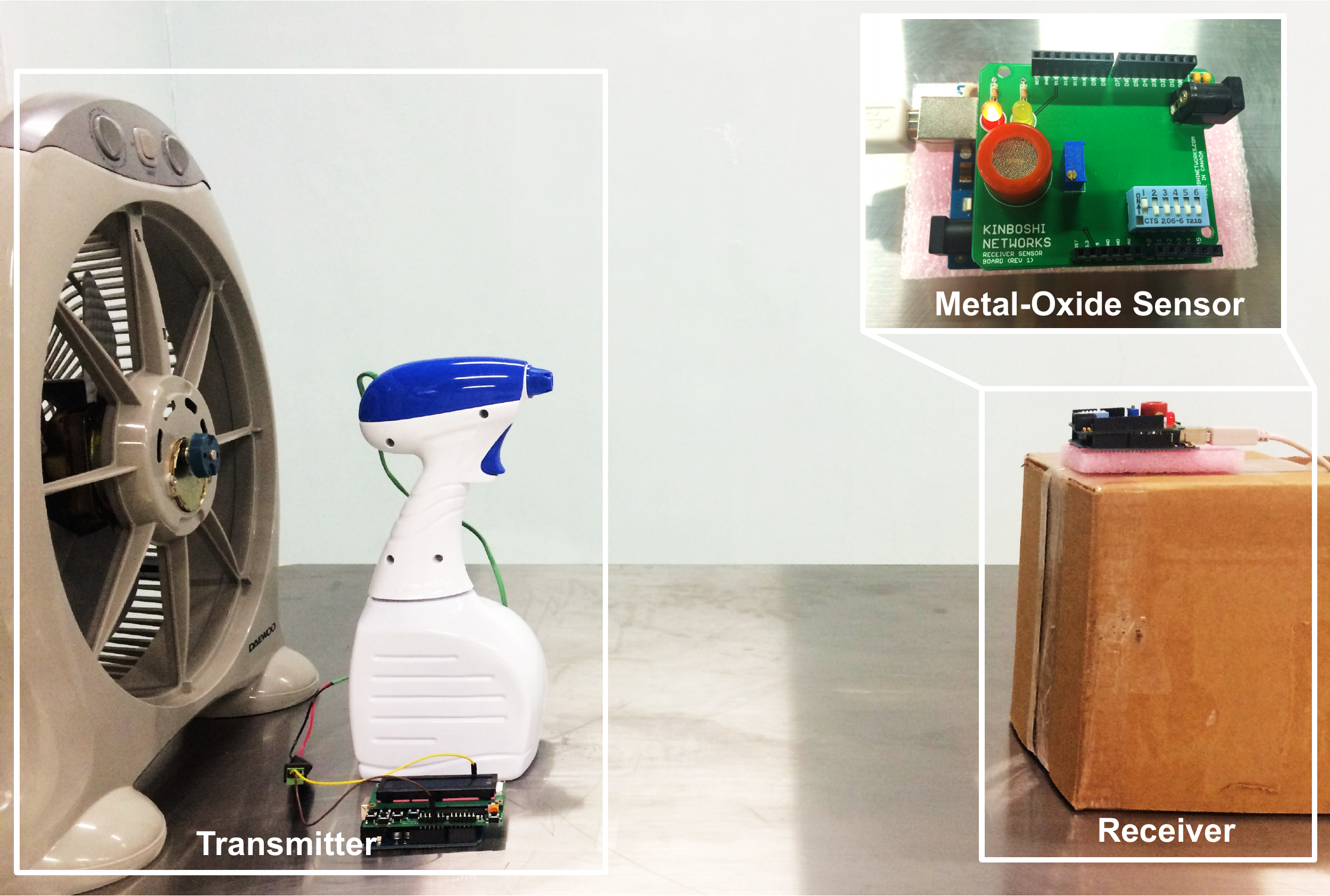}}}
  \caption{The tabletop platform with the MQ-3 metal-oxide sensor.}
  \label{Fig:testbed}
\end{figure}

The first experimental platform for molecular communication was presented in~\cite{far13}. This system, which was designed to be inexpensive and flexible, proved the feasibility of molecular communication, especially on the macroscale. This testbed has an electronically controllable spray that acts as a transmitter, a metal-oxide gas sensor that acts as a receiver, and a fan to assist the propagation. Two Arduino microcontrollers are also used to control the actions of the transmitter and the receiver. Fig.~\ref{Fig:testbed} highlights this system and its different components. 

One of the main advantages of molecular communication is its multi-scale property: molecular communication can be used on both the microscale and macroscale. Although the tabletop platform designed in~\cite{far13} is a macroscale system, some of the components can be shrunk to micro- and nanoscale. For example, metal-oxide sensors can be easily shrunk to a nanoscale, and be used for detection of different biological compounds~\cite{rah10, sol11}. Therefore, understanding how these sensors work can be very beneficial for both macroscale and microscale molecular communication.

As it was shown in~\cite{far14JSAC}, the system response of the testbed departs from the previous theoretical channel models used in the literature. Moreover, it was shown that the system tends to be nonlinear, where the nonlinearity can be modeled as noise. In~\cite{far14JSAC}, we attempted to add correction terms to the theoretical channel models. It was demonstrated that the proposed corrections to the system response would agree with experimental results. That work, however, considered only a single separation distance between the transmitter and the receiver and a single spray duration.

\begin{figure}[t]
 \centerline{\resizebox{\columnwidth}{!}{\includegraphics{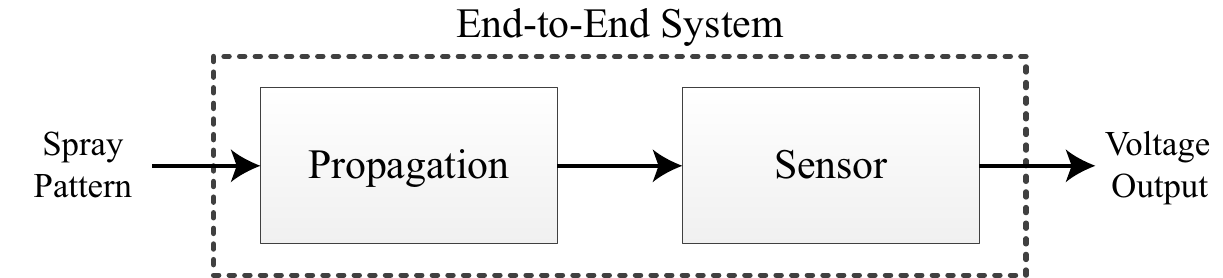}}}
  \caption{Block diagram of an end-to-end system.}
  \label{Fig:system}
\end{figure}

In this work, we build on our previous work and derive a more realistic analytical model for the testbed. The following are the new contributions in this work (compared to \cite{far14JSAC}).
\begin{itemize}
\item We consider a system model where the sensor and particle propagation are two separate systems. Previous work has modeled the tabletop platform as one system. By using two separate systems connected in series, a more realistic model is achieved.
\item We consider the end-to-end system response and how it changes with different system parameters such as distance, spray duration, and initial voltage (i.e., the initial concentration of the messenger molecules in the environment). Our previous work has considered a system model based only on a fixed set of system parameters.
\item A newly defined additive noise model is also presented for this systems, and is analyzed.
\end{itemize}

The rest of the paper is organized as follows. Section~\ref{Sec:System} defines the system models, and Section~\ref{Sec:UnivModel} analyzes the effect of the system parameters on channel response. A universal channel model is established in Section~\ref{Sec:gen}, and Section~\ref{Sec:Conc} concludes the paper.

\section{System Model}
\label{Sec:System}
As shown in Fig.~\ref{Fig:testbed} the transmitter is a spray, and the receiver is a metal-oxide sensor. A fan is used to assist the propagation, and the isopropyl alcohol is used to carry information. The end-to-end system block diagram of the tabletop platform is presented in Fig.~\ref{Fig:system}. Essentially the system input is a spray pattern, and the system output is the voltage readings. The system has two components: the propagation module, where the sprayed alcohol is carried to the sensor, and the sensor module, where the detected alcohol concentration is converted to voltage.

An impulse input is generated through a very short and instantaneous spray (e.g., 100ms). The system response to the impulse input is the impulse response of the system. In our previous work~\cite{far14JSAC}, the end-to-end system response of the system was based only on the chemical propagation model. Specifically, the model considered was
\begin{align}
\label{Eq:adsorb}
C(t) &= \frac{a}{\sqrt{t^3}}\exp{\left(-b\frac{(d-ct)^2}{t}\right)} 
\end{align}
where $a$, $b$, and $c$ are correction factors introduced to match the theoretical and experimental results, and $d$ is the separation distance between the transmitter and the receiver.

In this work, we consider a more accurate model by incorporating the sensor block as well.

\subsection{Metal-Oxide Sensors}
\label{Sec:Sensors}
The sensor used in the platform is the MQ-3 metal-oxide sensor. Metal-oxide sensors are cheap and can detect various gasses as well as volatile liquids such as alcohol~\cite{boc10}. They can also be shrunk to nanoscale and be used to detect various biological compounds~\cite{rah10 , sol11}. Moreover, metal-oxide sensors can be used to detect different chemicals~\cite{kim07MetOx}.

\begin{figure}[t]
 \centerline{\resizebox{.8\columnwidth}{!}{\includegraphics{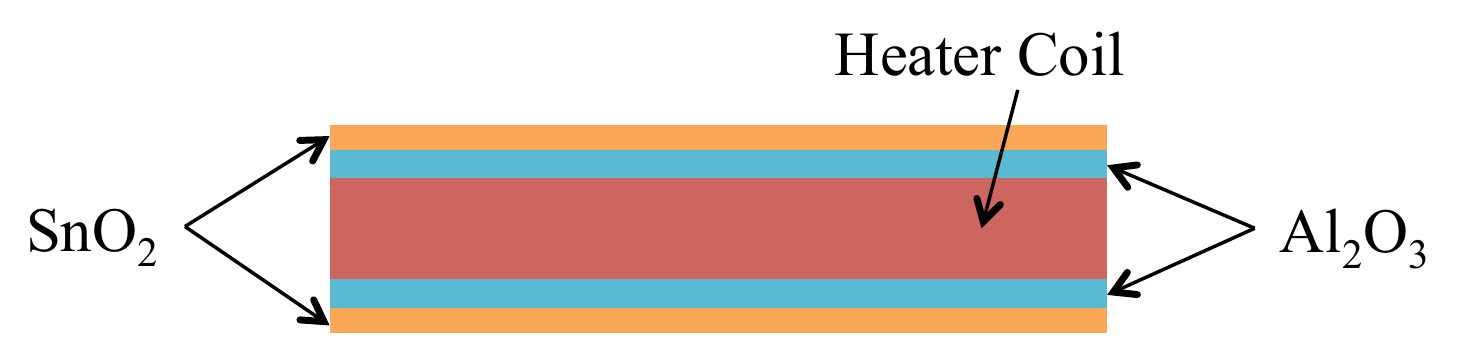}}}
  \caption{The sensor anatomy.}
  \label{Fig:sensDesign}
\end{figure}

The MQ-3 is a thin film tin dioxide (SnO$_2$) sensor, where an aluminium oxide (Al$_2$O$_3$) tube is covered with a thin layer of SnO$_2$. There is also a heater coil inside the aluminium oxide made from nickel–chromium alloy, which is used for resistive heating.  Fig.~\ref{Fig:sensDesign} shows the anatomy of the MQ-3 sensor.

The sensor works as follows. First, the heater coil heats up the sensing layer (i.e., the SnO$_2$ layer). According to the sensor's user manual,  to achieve the optimal sensor sensitivity, the sensor must be heated 24 to 48 hours prior to use. When the SnO$_2$ layer heats up, it becomes a semiconductor. When alcohol vapour approaches the sensor, it will go through an oxidization reaction that will in turn change the resistance of the  SnO$_2$ sensing layer~\cite{koh89}. In particular, the resistance of the SnO$_2$ sensing layer decreases as the concentration of alcohol increases in the vicinity of the sensor.

There is a well-known relationship between the resistance of a metal-oxide sensor and the concentration of target molecules~\cite{yam08}
\begin{align}
R_s &\cong a_1 C^n  
 \label{Eq:resistance}
\end{align}
where $R_s$ is the resistance of the sensor; $C$ is the concentration of molecules, and $a_1$ and $n$ are constants.
In~(\ref{Eq:resistance}), the constant, $a_1$, is empirically chosen, and the power law exponent, $n$, is specific to the type of target molecule. It can also be calculated theoretically from the reduced depletion depth and reduced reactivity~\cite{yam08}.

\begin{figure}[t]
 \centerline{\resizebox{\columnwidth}{!}{\includegraphics{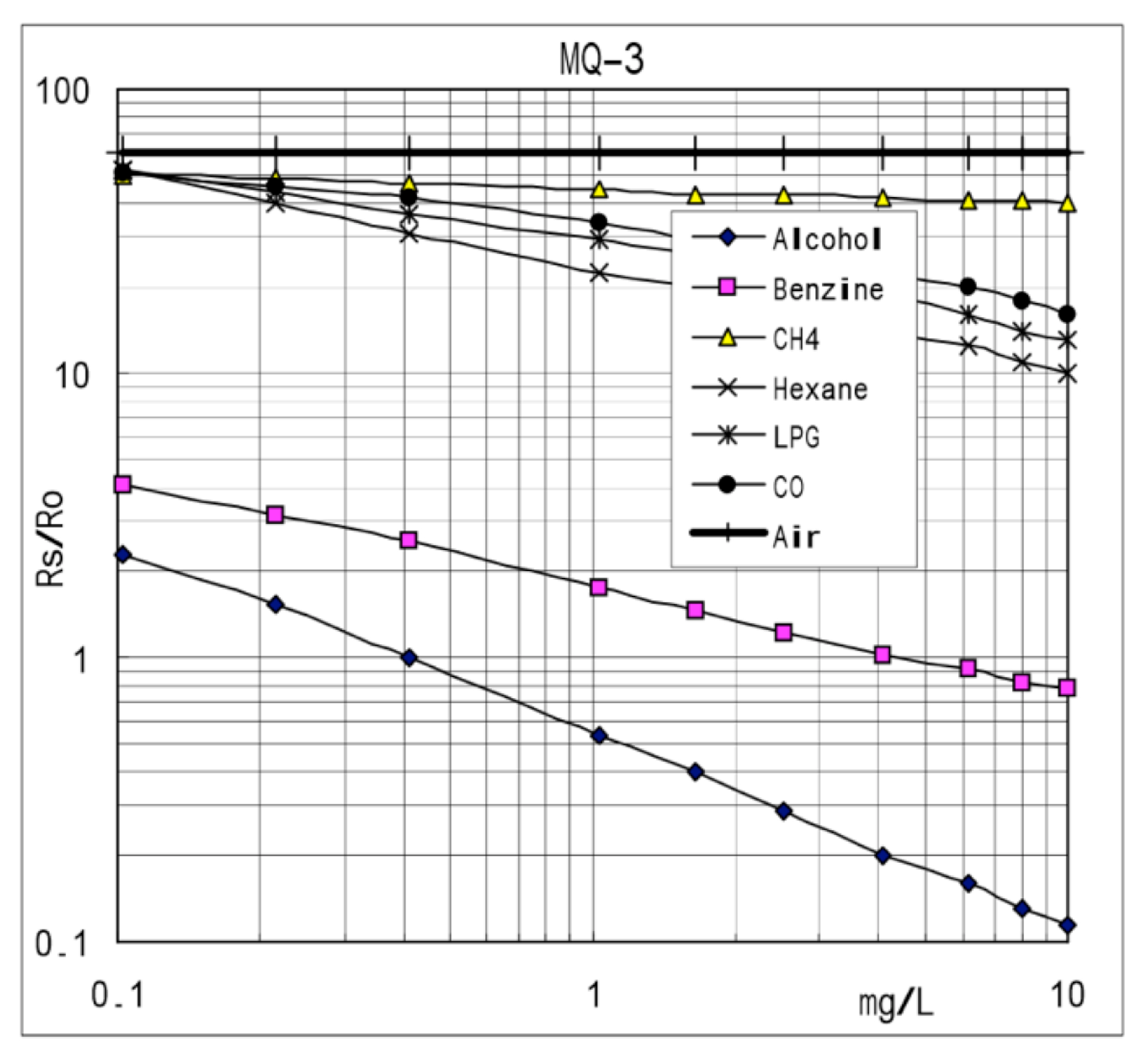}}}
  \caption{The sensor sensitivity plot from the datasheet~\cite{yam08}.}
  \label{Fig:senSenPlot}
\end{figure}
To estimate the constant $n$ provided in~(\ref{Eq:resistance}) for the MQ-3 sensor, we use the sensitivity graph in the MQ-3 data sheet. In this graph, shown in Fig.~\ref{Fig:senSenPlot}, the ratio $R_s/R_0$ is plotted against the concentration of different target gases, where $R_0$ is the resistance of the sensor at a specific concentration of the target gas $C_0$. In this plot, both axes are in log scale. Therefore we have
\begin{align}
\label{Eq:resistance2}\log \left( \frac{R_s(t)}{R_0} \right) &\cong \log \left( a_2 C(t)^n \right),
\end{align}
where $a_2=a_1/R_0$ is a new constant. From (\ref{Eq:resistance2}) it is clear that $n$ is the slope of the line corresponding to alcohol in Fig~\ref{Fig:senSenPlot}. Therefore, we estimate the value of $n=-0.65$, and the sensor resistance is related to the concentration with  
\begin{align}
\label{Eq:resisFinal}R_s(t) = a_1 C(t)^{-0.65},
\end{align}
where $a_1$ is an unknown constant.

\subsection{End-to-End Response}

As shown in Fig.~\ref{Fig:system}, the system consists of propagation and sensor detection. For the propagation phase, the expected system response is given by~\cite{Chae_CL}:
\begin{align}
\label{Eq:propRes}C(t) = M\frac{d}{\sqrt{4\pi D t^3}} \exp \left(-\frac{(d-vt)^2}{4Dt} \right),
\end{align}
where $M$ is the number of alcohol molecules released by the short spray; $d$ is the distance between the transmitter and the receiver; $D$ is the effective diffusion coefficient, and $v$ is the average velocity of the wind flow. Substituting (\ref{Eq:propRes}) into (\ref{Eq:resisFinal}), the end-to-end system response is given by
\begin{align}
\label{Eq:EtoEres}R_s(t) = a_1 \left[M\frac{d}{\sqrt{4\pi D t^3}} \exp \left(-\frac{(d-vt)^2}{4Dt} \right) \right] ^{-0.65}.
\end{align}
The change in sensor resistance is measured using a simple voltage divider circuit. 

Since $M$, $a_1$, the effective diffusion coefficient $D$, and the average velocity $v$ cannot be measured accurately, the end-to-end system response of the system is given by 
\begin{align}
h(t;a,b,c)=a\left[\frac{d}{\sqrt{4\pi bt^3}}\exp\Big(-\frac{(d-ct)^2}{4bt}\Big)\right]^{-0.65}
\label{Eq:M}
\end{align}
where $a$, $b$, $c$ are unknown coefficients, and $d$ is the distance between the transmitter and the receiver. Specifically, $a$ is a the product of constant $a_1$ and the number of transmitted molecules $M$; $b$ is the effective diffusion, and $c$ is the average velocity. By obtaining the proper values for the coefficients, a generalized universal model can be established. In the next section, we use experimental data to estimate the values of these three coefficients, and observe how they change with respect to different system parameters.
 
 \begin{table}[!t]
\caption{System Parameters.}
\begin{center}
\begin{tabular}{|c|c|c|c|c|}
\hline
Distance (m) & 2  & 3  & 4  & 5 \\ \hline
Spraying Duration (ms) & 50  & 100  & 150  & 200 \\ \hline
Initial Voltage (V) & 1.0  & 1.3 & 1.6 & 1.9\\ \hline
\end{tabular}
\end{center}
\label{tb:parameters}
\end{table}

 \begin{figure*}[!t]
  \centerline{\resizebox{2.7\columnwidth}{!}{\includegraphics{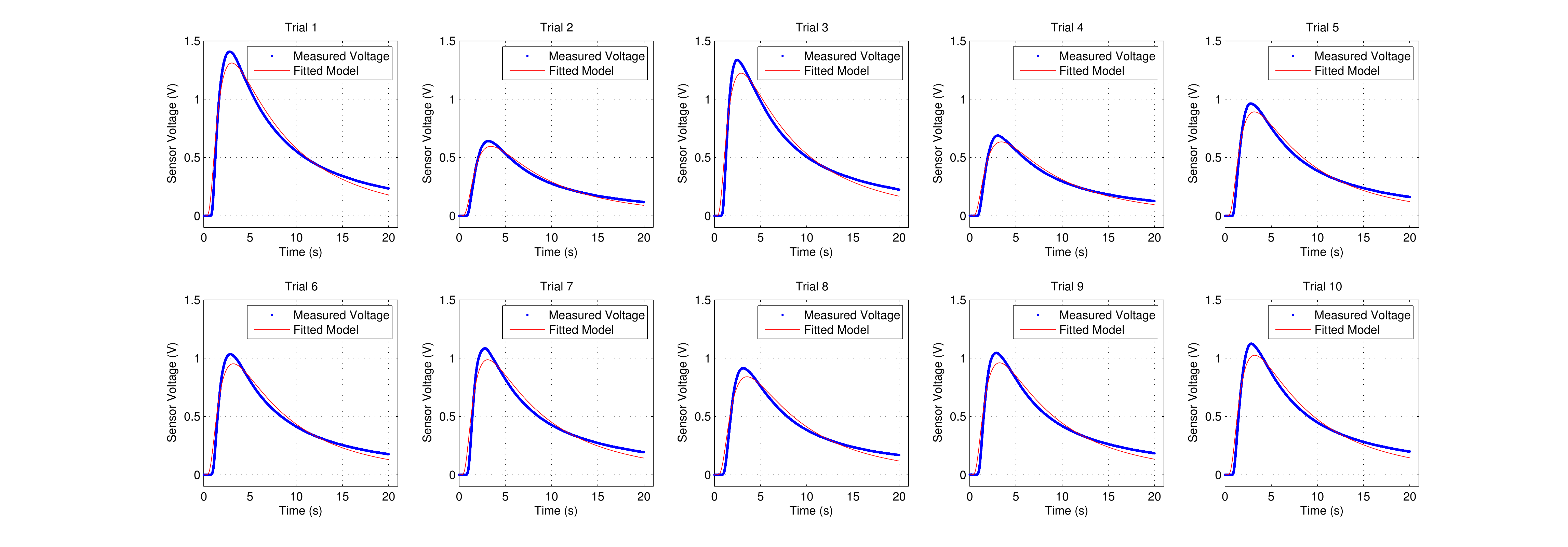}}}
   \caption{One fitting example with three unknown coefficients when system parameters are 2~m, 150~ms, and 1.3~V.}
   \label{Fig:fitting1}
 \end{figure*}
\begin{figure*}[!t]
        \centering
        \begin{subfigure}[b]{0.31\textwidth}
                \centering
                \includegraphics[width=\textwidth]{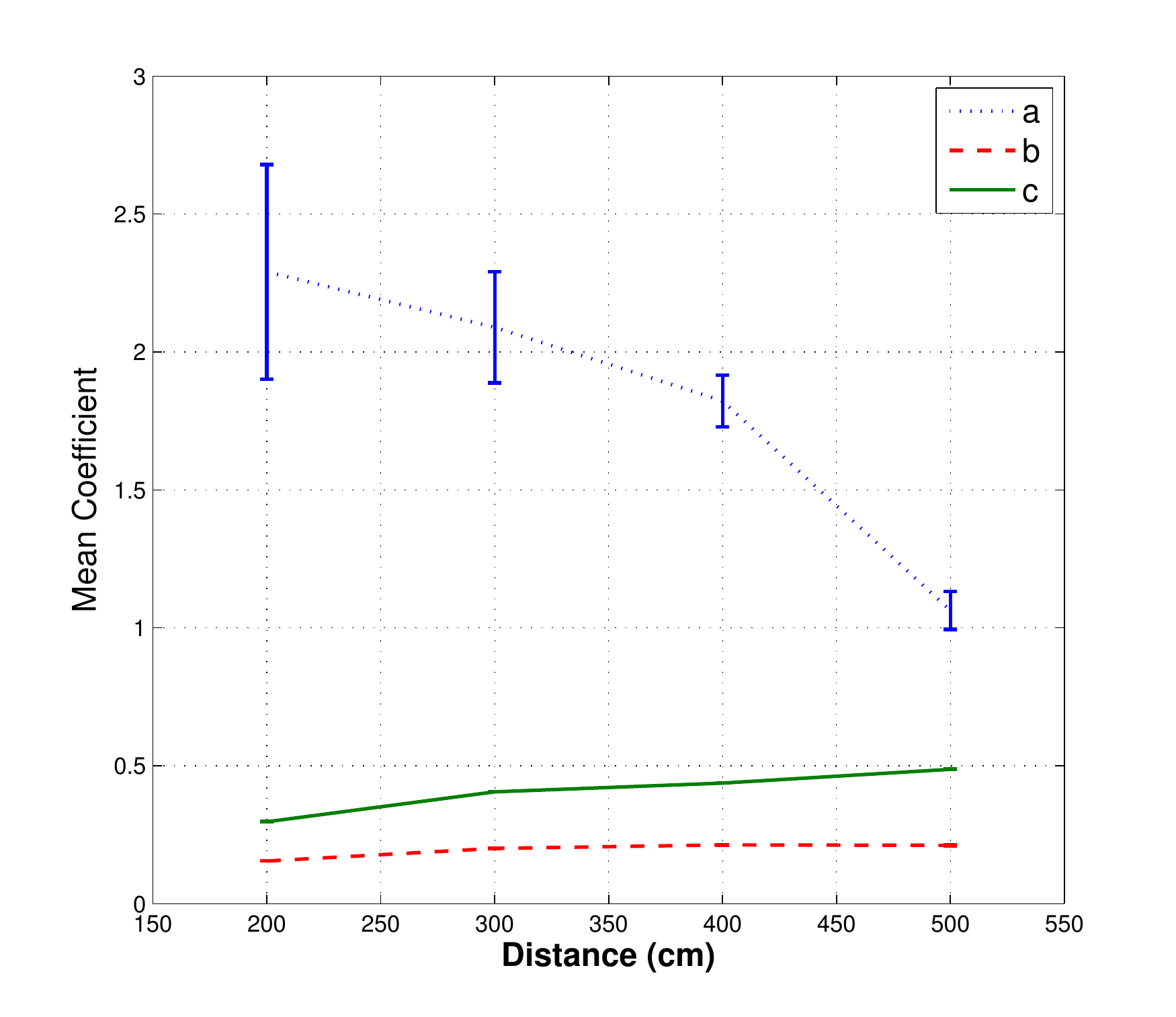}
                \caption{Distance effect on the coefficients.}
                \label{Fig:}
        \end{subfigure}%
       ~
        \begin{subfigure}[b]{0.31\textwidth}
                \centering
                \includegraphics[width=\textwidth]{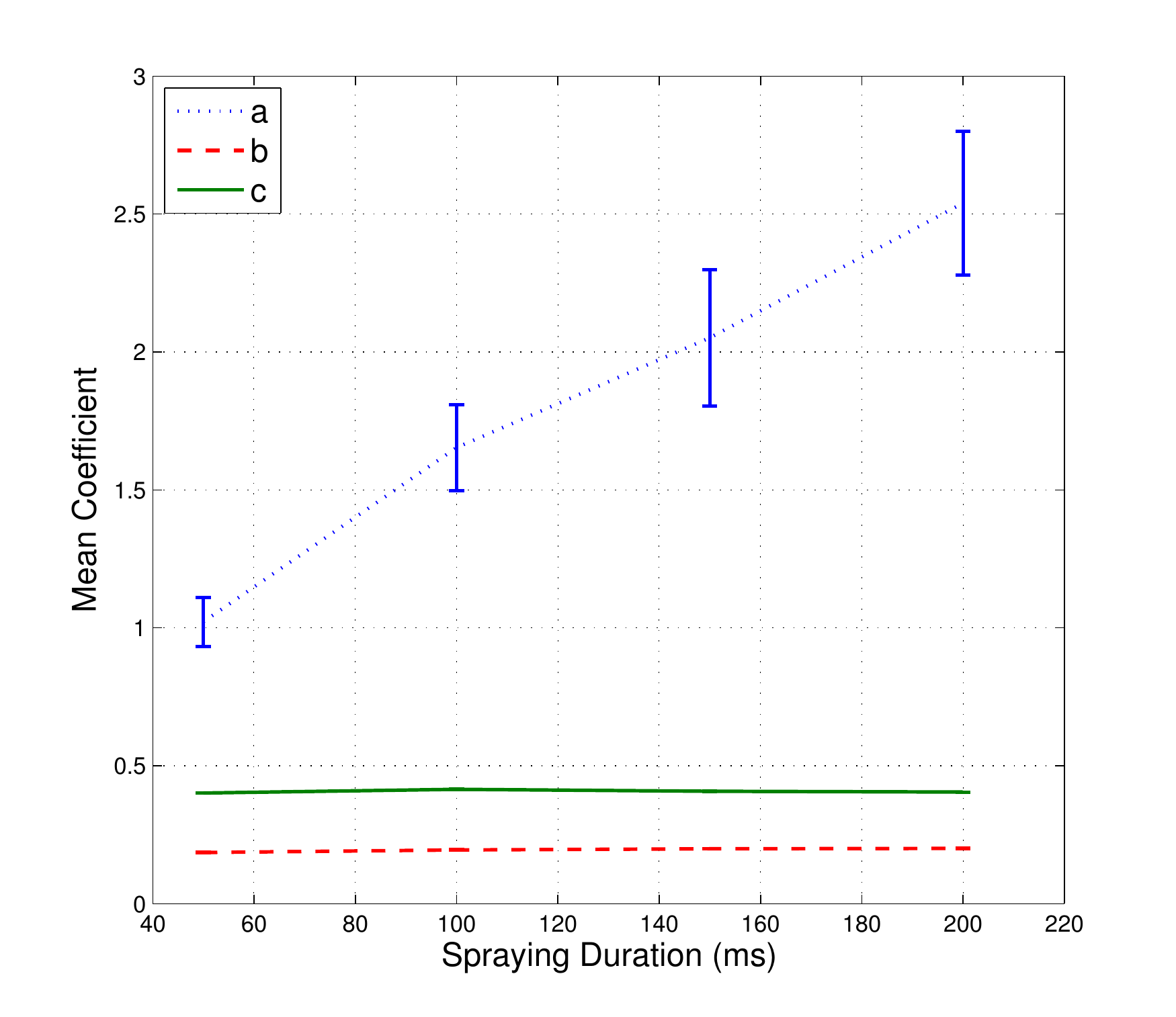}
                \caption{Spraying Duration effect on the coefficients.}
                \label{Fig:}
        \end{subfigure}
                ~
        \begin{subfigure}[b]{0.31\textwidth}
                \centering
                \includegraphics[width=\textwidth]{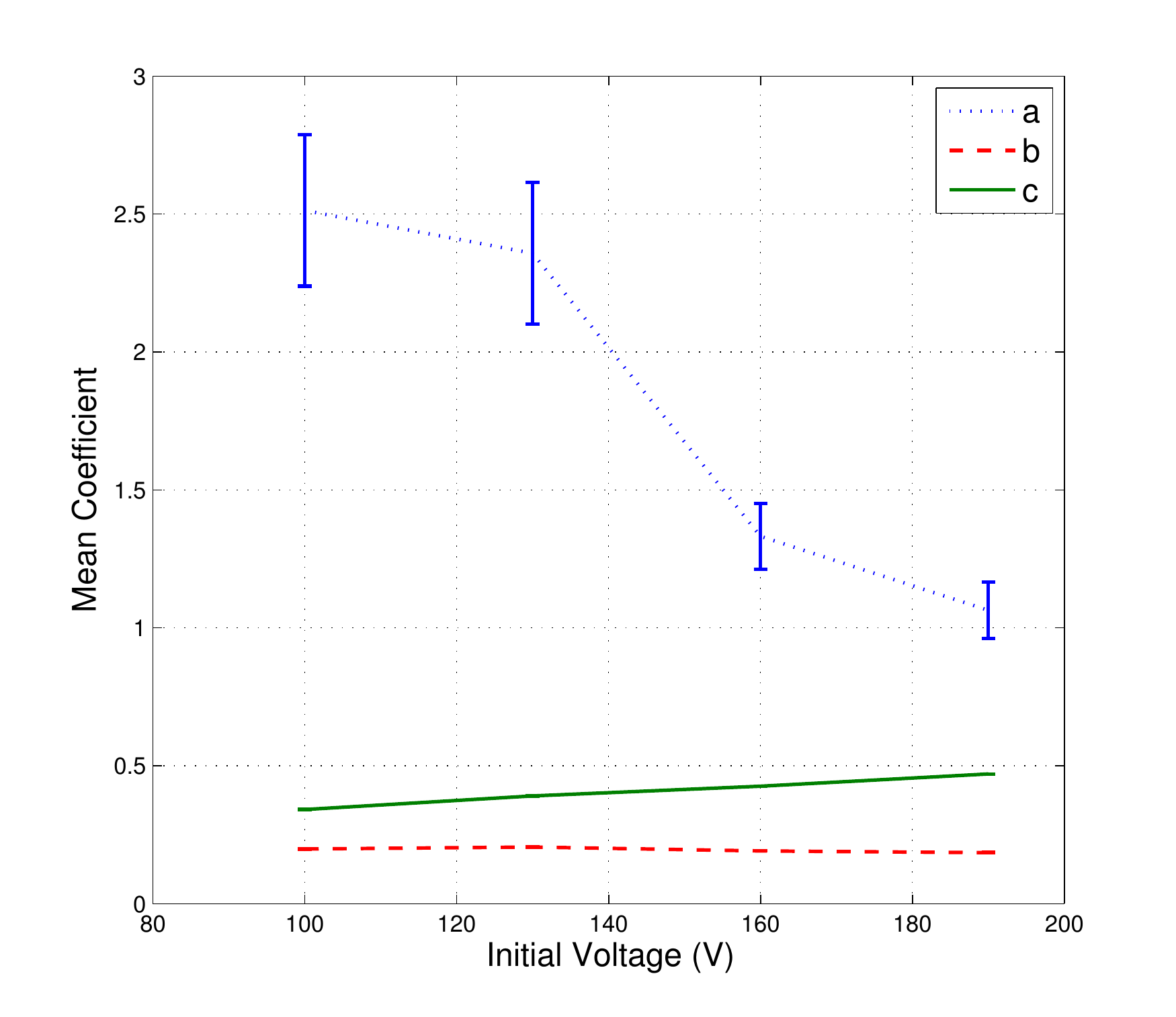}
                \caption{Initial Voltage effect on the coefficients.}
                \label{Fig:}
        \end{subfigure}
        ~ 
        \caption{Averaged effect of the system parameters.}\label{Fig:Effect_coef1}
\end{figure*}

\section{Universal Model Based on Experiments}
\label{Sec:UnivModel}
To find the value of the unknown coefficients as a function of system parameters, the tabletop platform shown in Fig.~\ref{Fig:testbed} is used to measure the end-to-end impulse response of the system using, as messenger molecules, isopropyl alcohol. 
To see the effects of different system parameters on the channel responses, three system parameters are considered: distance between the transmitter and the receiver, spraying duration (which is related to the number of the transmitted molecules), and initial voltage or initial concentration of the messenger molecules in the environment. Each parameter is varied over four different values as shown in Table~\ref{tb:parameters}. Therefore, there are 4$\times$4$\times$4$=$64 different cases, and the end-to-end impulse response of each case is measured during 10 different trials. Thus, to collect all the data to estimate the unknown coefficients, in this paper, we carry out, in total, 640 sets of experimental trials. 
\begin{figure*}[!t]
        \centering
        \begin{subfigure}[b]{0.31\textwidth}
                \centering
                \includegraphics[width=\textwidth]{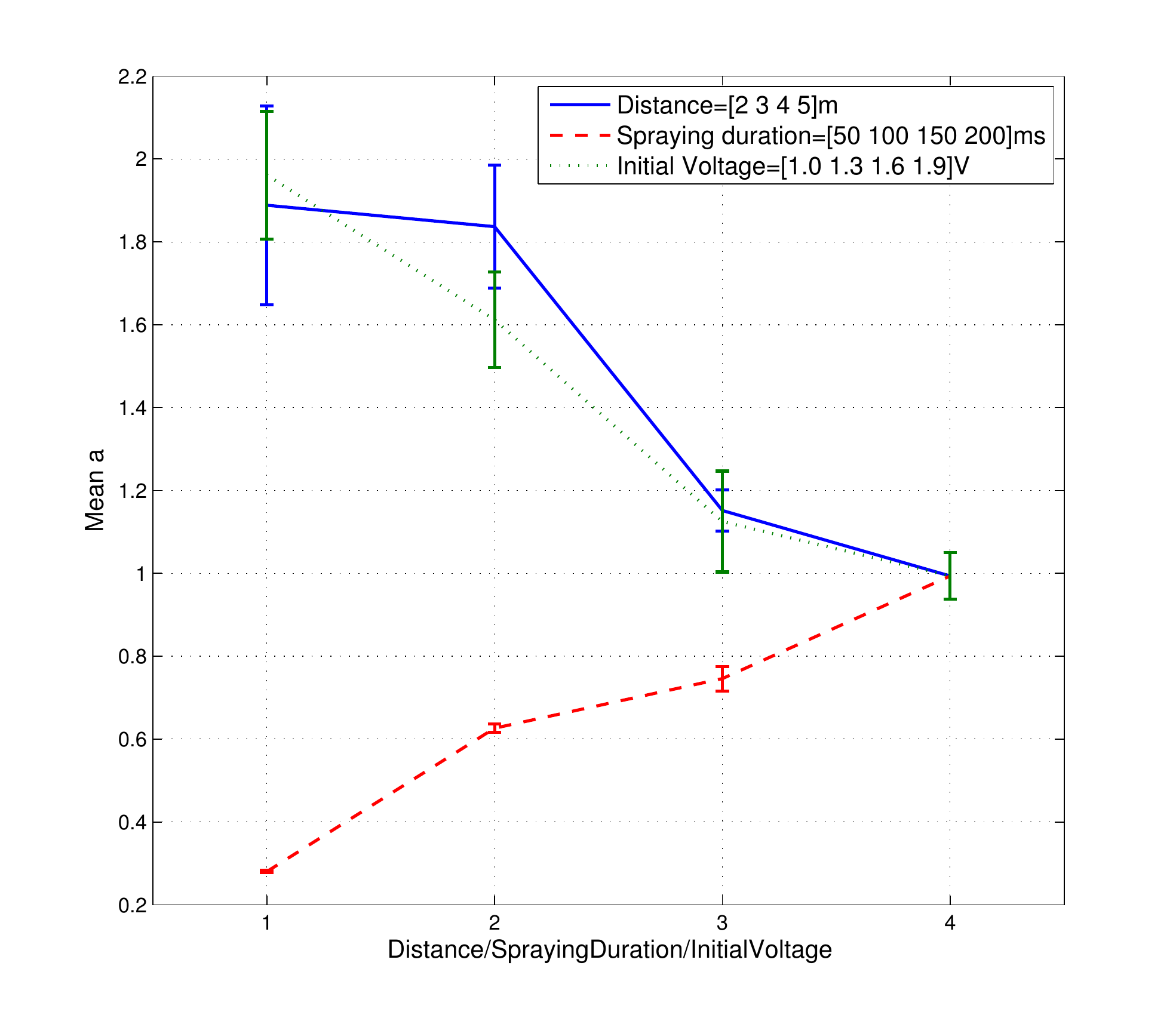}
                \caption{The averaged effect of system parameters on coefficient $a$.}
                \label{Fig:Effect_coef2}
        \end{subfigure}%
       ~
        \begin{subfigure}[b]{0.31\textwidth}
                \centering
                \includegraphics[width=\textwidth]{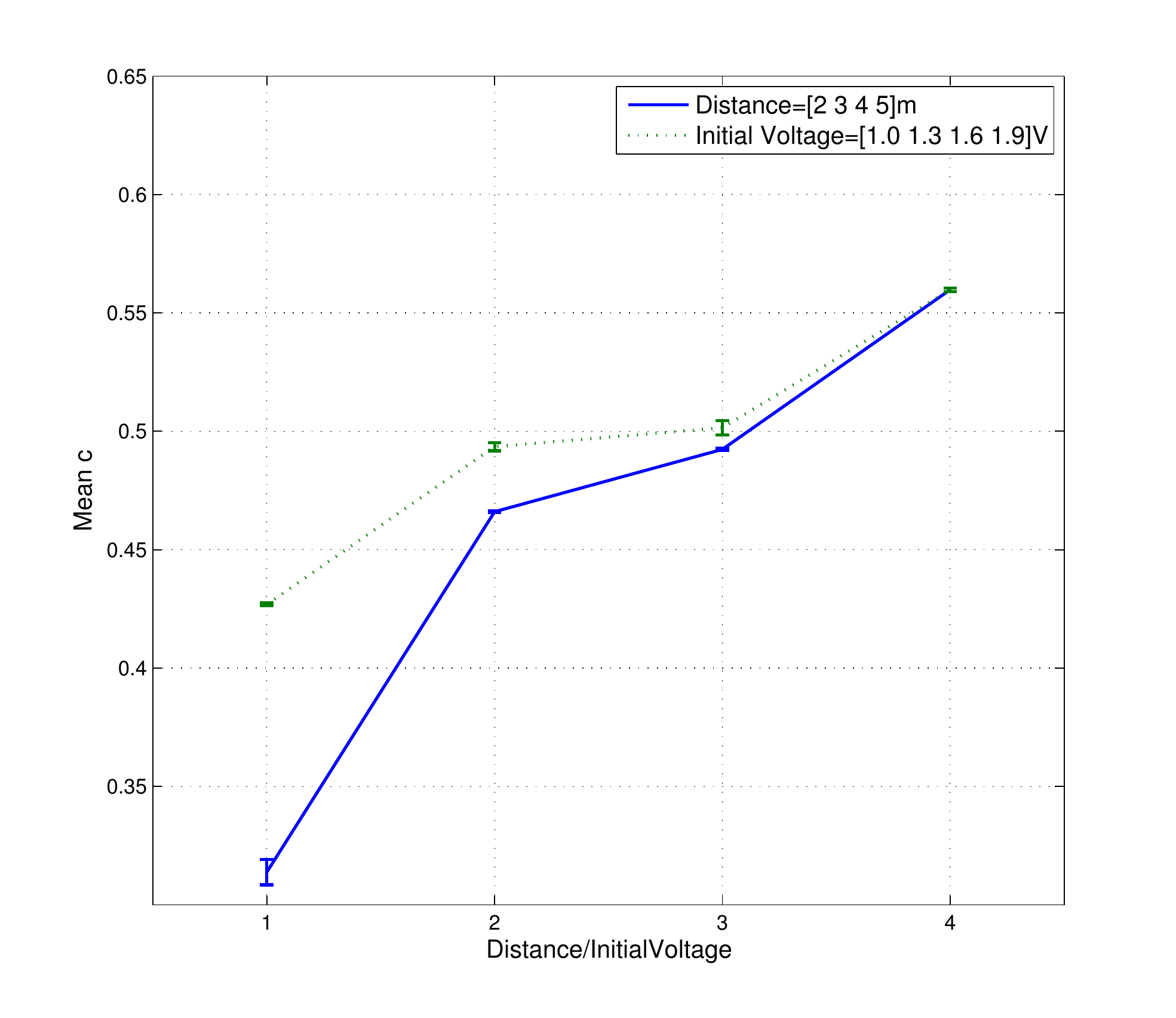}
                \caption{The averaged effect of system parameters on coefficient $c$.}
                \label{Fig:Effect_coef3}
        \end{subfigure}
                ~
        \begin{subfigure}[b]{0.31\textwidth}
                \centering
                \includegraphics[width=\textwidth]{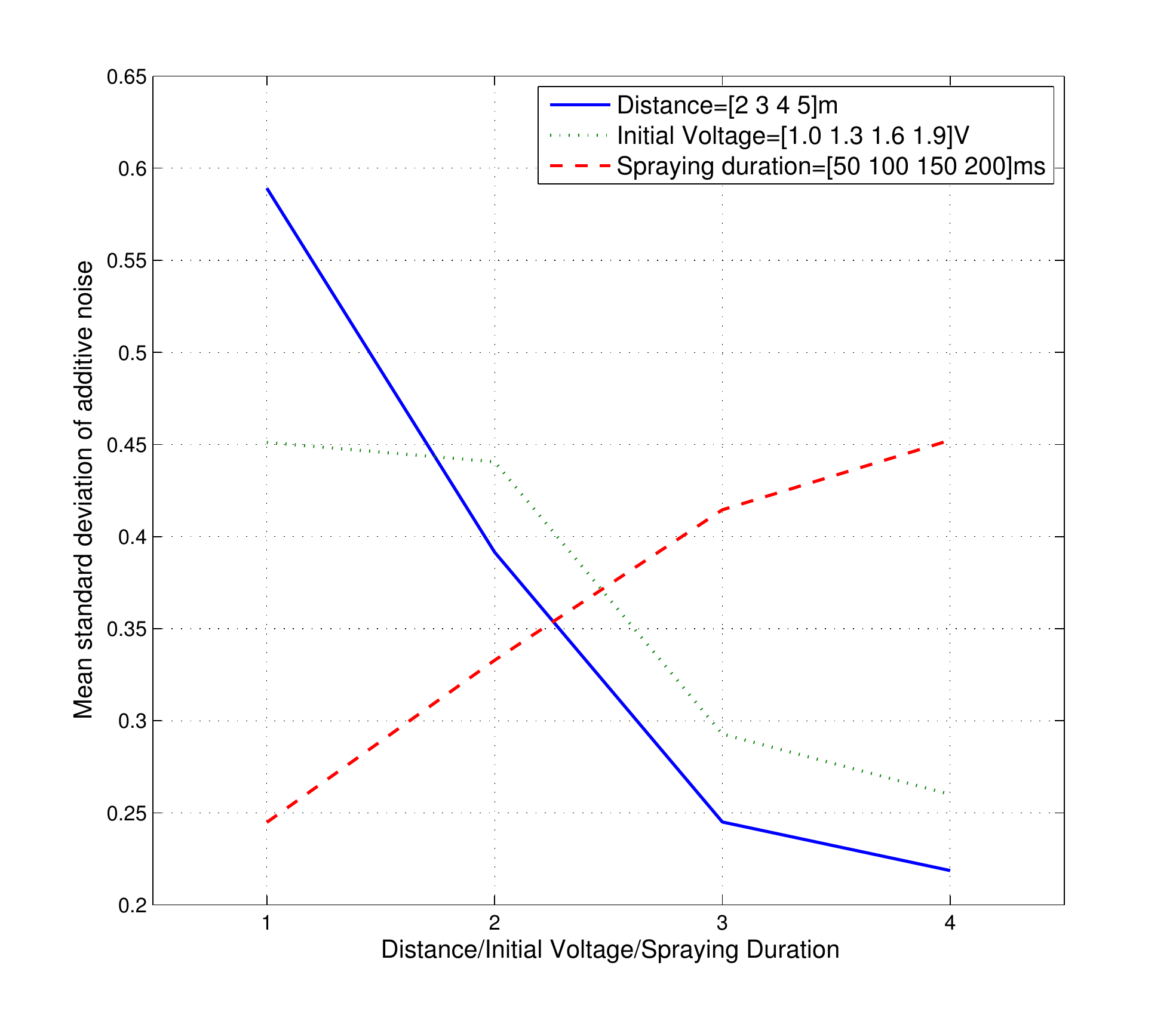}
                \caption{The averaged effect of system parameters on the standard deviation of the additive noise, $\sigma$.}
                \label{Fig:Effect_coef4}
        \end{subfigure}
        ~ 
        \caption{The averaged effect of system parameters to the coefficients, $a$ and $c$, and standard deviation of the additive noise, $\sigma$. The x-axis indicates the index of the system parameter vectors as represented in the legend.}\label{Fig:Effect_coef}
\end{figure*}

\subsection{Coefficients}
\label{Subsec:coef}

To estimate the value of the unknown coefficients in the analytical end-to-end model (\ref{Eq:M}), we use a nonlinear least squares curve fitting technique used in our previous work \cite{far14JSAC}. Assuming that there are $N$ points in the sensor measurement from each trial, then  (\ref{Eq:M}) can be discretized as $h(t_k, \mathbf{p})$ ($k \in \{1, 2,\cdots, N\}$), where $t_k$ are sampling times, and $\mathbf{p}=[a,b,c]^T$ is the parameter vector representing the three unknown constants. The coefficient estimation problem can be formulated as
\begin{align}
	 \argmin_{\mathbf{p}} \sum_{k=1}^N \big( O(t_k)-h(t_k,\mathbf{p}) \big)^2
\end{align}
where $O(t_k)$ are the experimental observations.

The coefficients are estimated for all the 640 data sets using the least squares estimation technique. For example, the fitted results are shown in Fig.~\ref{Fig:fitting1} for the case when the distance between the transmitter and the receiver is 2~m, the spray duration is 150~ms, and the initial voltage is 1.3~volts. It is apparent that our model function fits very well to the measurement data. The root mean square error (RMSE) is 0.0484, which is a bit smaller than with the previous model.

Fig.~\ref{Fig:Effect_coef1} shows the effect of the system parameters on the coefficients. Each point in the plot represents the average value of the corresponding coefficient across 160 trials. The standard deviation bars represent the standard deviation across trials. As can be seen, coefficient $a$ changes frequently with changing system parameters, while coefficient $c$ varies slightly with respect to system parameters, and $b$ is mostly a constant. Specifically, coefficient $a$ changes with distance, spray duration, and initial voltage; $c$ changes only with distance and initial 
voltage (not with spraying duration). These results agree with the physical characteristics of the system. For example, the effective diffusion coefficient is expected to be a constant.  From Fig.~\ref{Fig:Effect_coef1}(b), we can see that as the spray duration increases, $a$ increases linearly and the variance of $a$ also increases. This is because more molecules are released as the spray duration increases thereby increasing M in~(\ref{Eq:EtoEres}).

Based on these observations, we assume $b$ to be constant and be equal to the average values of $b$ over 640 trials. Note that since the variance of $b$ is so small, this is a very good estimation for this coefficient. Let $b^*$ be this average value with $b^*=0.1950$. 

When we perform least squares curve fitting with the fixed coefficient $b^*$ and unknown coefficients $a$ and $c$, the fitted curves are as good as the ones shown in Fig.~\ref{Fig:fitting1}. Therefore, we leave out those plots to avoid duplications. 
Fig.~\ref{Fig:Effect_coef}(a) and \ref{Fig:Effect_coef}(b) show the behaviour of the coefficients with the fixed $b^*$. As can be seen, both $a$ and $c$ change with respect to the system parameters. Let $d$, $s$, and $\nu$ be the distance, spraying duration, and the initial voltage, respectively. Then from (\ref{Eq:M}) we have 
\begin{align}
\begin{split}
& h(t;a,b*,c)=a\Bigg[\frac{d}{\sqrt{4\pi b^*t^3}}\exp\Big(-\frac{(d-ct)^2}{4b^*t}\Big)\Bigg]^{-0.65},\\
& a = f(d,s,\nu),~c = g(d,\nu),\\
& h(t;d,s,\nu) = \\
&f(d,s,\nu) \Bigg[\frac{d}{\sqrt{4\pi b^*t^3}}\exp\Big(-\frac{(d-g(d,\nu)t)^2}{4b^*t}\Big)\Bigg]^{-0.65}.
\end{split}\label{Eq:f}
\end{align} 
where $h(t;d,s,\nu)$ is a function of time, $t$, with three coefficients.

\subsection{Noise}
\label{Subsec:noise}
An important part of the system is the random nature of the end-to-end impulse response. Some of the factors that contribute to the randomness are: the spray, which is not precise enough to spray the same amount of alcohol across trials, the random propagation due to diffusion and turbulent flows, and other phenomena. Therefore, to obtain a complete system model, the random effect must be represented as noise. To establish a noise model, an additive noise is defined as the difference between the experimental observation and the model function shown below. 
\begin{align}
N_i(t) = O_i(t) - h(t;d,s,\nu)
\label{Eq:addnoise}
\end{align}
where $N_i(t)$ represents the additive noise and $O_i(t)$ represents the observation or measurement for each trial $i$ ($i=1,2,~10$ for a specific distance, spray duration, and initial voltage). From Fig.~\ref{Fig:Effect_coef}(a) and \ref{Fig:Effect_coef}(b), we can see that for coefficient c, the standard deviation across the trials is slight. Therefore, since only coefficient $a$ has a large variance across different trials we have 
\begin{align}
&O_i(t) = \nonumber \\ 
&(f(d,s,\nu) + N)\Bigg[\frac{d}{\sqrt{4\pi b^*t^3}}\exp\Big(-\frac{(d-g(d,\nu)t)^2}{4b^*t}\Big)\Bigg]^{-0.65},
\label{Eq:addnoise2}
\end{align}
where $N$ is the noise introduced by the system, with
\begin{align}
N_i(t) &=  N\Bigg[\frac{d}{\sqrt{4\pi b^*t^3}}\exp\Big(-\frac{(d-g(d,\nu)t)^2}{4b^*t}\Big)\Bigg]^{-0.65}.
\label{Eq:addnoise3}
\end{align}
Let $a_{O_i}$ be the best least squares fit for coefficient $a$ for the trial experimental observation $O_i(t)$. Then a noise sample for $N$ is obtained by:
\begin{align}
N &=  a_{O_i} - f(d,s,\nu)
\label{Eq:addnoise4}
\end{align}
Since $f(d,s,\nu)$ is obtained from the average value of coefficient $a$ over different trials, the mean of $N$ becomes zero for all cases. As represented in Fig.~\ref{Fig:Effect_coef}(c), however, the standard deviation of $N$ changes with system parameters showing  behaviour similar to $a$. Thus, it is also possible to express the standard deviation as a function of the system parameters as in analyzing coefficient $a$.
\begin{align}
\begin{split}
&E[N]=0\\
&\sqrt{Var[N]}=\sigma=L(d,s,\nu)
\end{split}\label{Eq:meanvar}
\end{align}
where $E[\cdot]$ and $Var[\cdot]$ denotes the expectation value and variance, respectively.

\section{A Complete End-to-End Model}
\label{Sec:gen}
In this section we derive a complete end-to-end model for the tabletop platform. From Section~\ref{Sec:UnivModel}, coefficient $a$, its additive noise's variance, and $c$ have a close relationship with the system parameters. Therefore, our goal is to estimate the functions $f(d,s,\nu)$, $g(d,\nu)$, and $L(d,s,\nu)$. 

\subsection{Estimating the Coefficient Functions}
From Figs.~\ref{Fig:Effect_coef}(a), \ref{Fig:Effect_coef}(b), and \ref{Fig:Effect_coef}(c), it can be seen that the coefficients $a$,  $c$ and the noise variance change relatively linearly with respect to the different system parameters. Therefore we estimate the functions $f(d,s,\nu)$, $g(d,\nu)$, and $L(d,s,\nu)$ as linear functions, respectively
\begin{align}
  f(d,s,\nu) &= \beta_d^{(f)} d+ \beta_s^{(f)} s + \beta_\nu^{(f)} \nu + \beta_0^{(f)}, \label{eq:linF}\\
  g(d,\nu)&= \beta_d^{(g)} d+ \beta_\nu^{(g)} \nu + \beta_0^{(g)}, \label{eq:linG} \\
  L(d,s,\nu)&=\beta_d^{(L)} d+ \beta_s^{(L)} s + \beta_\nu^{(L)} \nu + \beta_0^{(L)}, \label{eq:linL} 
\end{align}
where all the $\beta$ variables are constants. To estimate the values of the $\beta$ constants, we use again the linear least squares method. Table \ref{tb:para_afit} shows the estimated values of these constants.

\begin{table}[!h]
\caption{The estimated $\beta$ coefficients in (\ref{eq:linF})--(\ref{eq:linL}).}
\begin{center}
\begin{tabular}{|c|c|c|c|c|c|}
\hline
Coefficient 	& Value	   & Coefficient      & Value  & Coefficient 		& Value \\ \hline \hline
$\beta_d^{(f)}$ & -0.4188  & $\beta_d^{(g)}$  & 0.0709 & $\beta_d^{(L)}$ 	& -0.1258  \\ \hline
$\beta_s^{(f)}$ & 0.0098   & -				  &  -     & $\beta_s^{(L)}$	& 0.0014	\\ \hline
$\beta_\nu^{(f)}$ & -1.7873& $\beta_\nu^{(g)}$& 0.1362 & $\beta_\nu^{(L)}$ & -0.2403	\\ \hline
$\beta_0^{(f)}$ & 4.6469   & $\beta_0^{(g)}$  & -0.0427& $\beta_0^{(L)}$	& 0.9738	\\ \hline
\end{tabular}
\end{center}
\label{tb:para_afit}
\end{table}
\subsection{Verification}

\begin{figure}[!t]
 \centerline{\resizebox{1.1\columnwidth}{!}{\includegraphics{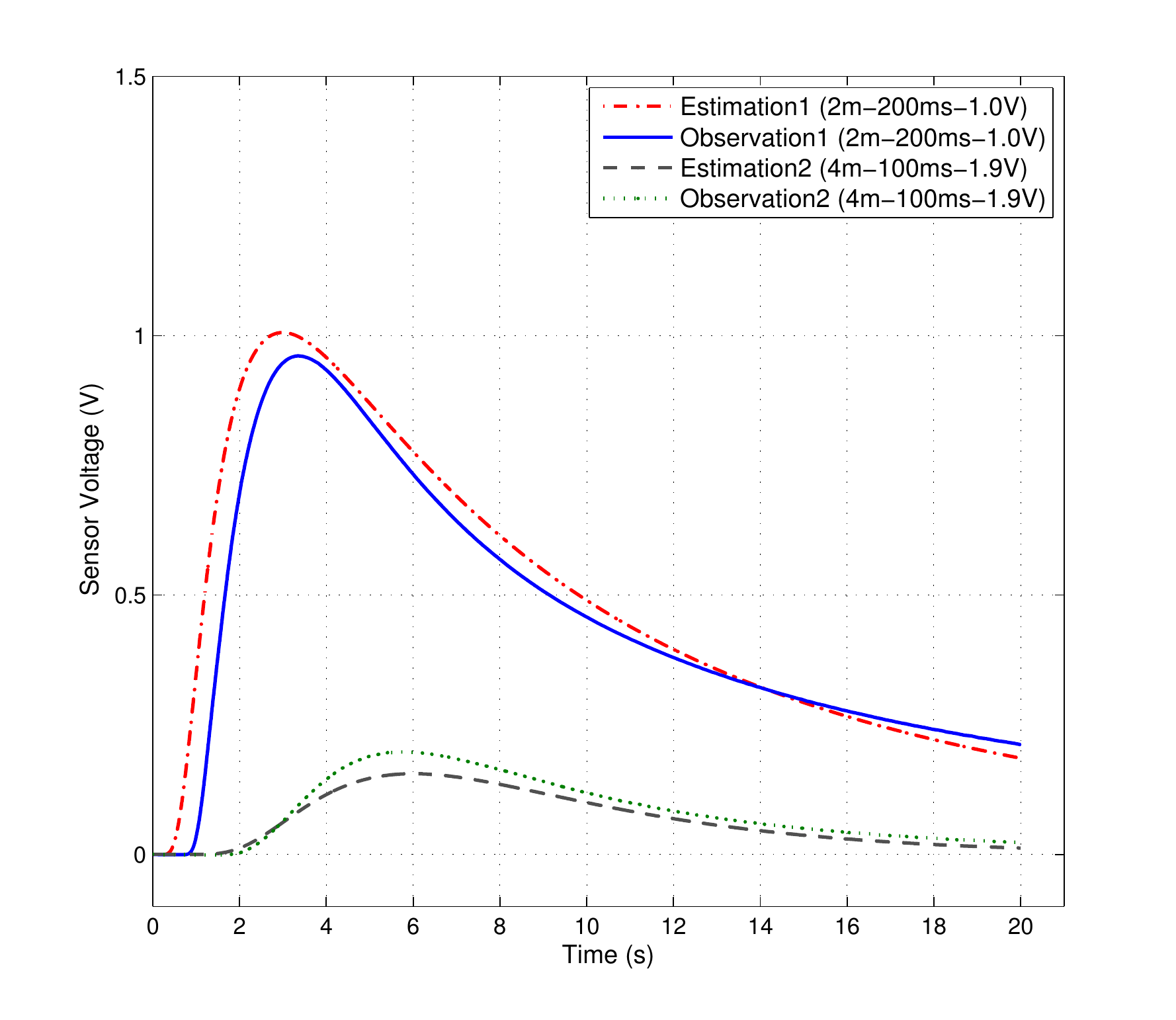}}}
  \caption{Comparison of the estimation from the obtained model and the experimental channel observations.}
  \label{Fig:test}
\end{figure}

To verify the estimation in the previous section, two data sets are selected: 1) a distance of 2.5~m, a spray duration of 130~ms, and an initial voltage of~1.5 V and 2) 3.5~m, 170~ms, and 1.1~V. 
Fig.~\ref{Fig:test} compares the comparison between the estimated system model and the averaged observations, and it shows that the two results match fairly well.
Thus, the newly developed end-to-end channel model can be utilized for whichever values of the distance, spraying duration, and initial voltage. 
%







\section{Conclusions}
\label{Sec:Conc}
In this work, we developed an end-to-end channel model based on experimental observations using the tabletop platform. Unlike previous work on the channel model, it is represented as two separate systems of the sensor and particle propagation making it a more realistic model. Also, it is generalized with three different system parameters: distance, spraying duration, and initial voltage. Furthermore, the randomness of the system is represented as an additive noise model. 
The two data sets also confirm that the newly developed universal channel model works well. 
For future work, we will propose new modulation techniques appropriate for this platform using this channel model, and compare the performances depending on the system parameters.  

\section*{Acknowledgement}
This research was in part supported by the MSIP (Ministry of Science, ICT $\&$ Future Planning), Korea, under the “IT Consilience Creative Program” (NIPA-2014-H0201-14-1002) supervised by the NIPA (National IT Industry Promotion Agency) and by the Basic Science Research Program (2014R1A1A1002186) funded by MSIP, Korea, through the National Research Foundation of Korea.

\bibliographystyle{IEEEbib}

\bibliography{MolCom_YearSorted}

\end{document}